\documentclass[aps,prd,a4paper,onecolumn,amsmath,showpacs,superscriptaddress,nofootinbib,preprintnumbers,notitlepage]{revtex4-1}
 
\usepackage{verbatim}
\usepackage[T1]{fontenc}
\usepackage[utf8]{inputenc}
\usepackage[american]{babel}
\usepackage{epsfig}
\usepackage{graphicx,subcaption,caption}
\usepackage{pgfplots}
\usepackage{booktabs}
\usepackage{multirow}
\usepackage{dcolumn}
\usepackage{amsmath}
\usepackage{mathtools}
\usepackage{amsfonts}
\usepackage{amssymb}
\usepackage{epstopdf}
\usepackage{bm}
\usepackage{siunitx}
\usepackage{braket}
\usepackage{enumitem}
\usepackage{soul}
\usepackage{ulem}
\usepackage{color}
\usepackage{transparent}
\usepackage{pifont}
\usepackage[font={small}]{caption}

\definecolor{navyblue}{rgb}{0.0, 0.0, 0.5}
\definecolor{royalblue}{rgb}{0.25, 0.41, 0.88}
\definecolor{cadmiumgreen}{rgb}{0.0, 0.42, 0.24}
\definecolor{blue-violet}{rgb}{0.54, 0.17, 0.89}
\definecolor{darkviolet}{rgb}{0.58, 0.0, 0.83}
\definecolor{orange(colorwheel)}{rgb}{1.0, 0.5, 0.0}

\usepackage{hyperref}
\hypersetup{
    colorlinks=true, 
    linkcolor=royalblue, 
    citecolor=magenta}

\newcommand\be{\begin{equation}}
\newcommand\ee{\end{equation}}

\newcommand\bea{\begin{eqnarray}}
\newcommand\eea{\end{eqnarray}}

\usepackage{booktabs}
\usepackage{multirow}
\usepackage{dcolumn}
\usepackage{colortbl}

\definecolor{magenta(process)}{rgb}{1.0, 0.0, 0.56}

\definecolor{darkspringgreen}{rgb}{0.09, 0.45, 0.27}

\definecolor{royalblue(web)}{rgb}{0.25, 0.41, 0.88}

\begin{document}

\title{RPS Thermodynamics of Taub-NUT AdS Black Holes in the Presence of Central Charge and the Weak Gravity Conjecture}

\author{Jafar Sadeghi}
\email{pouriya@ipm.ir}
\affiliation{Department of Physics, University of Mazandaran, P. O. Box 47416-95447, Babolsar, Iran}

\author{Mehdi Shokri}
\email{mehdishokriphysics@gmail.com}
\affiliation{School of Physics, Damghan University, P. O. Box 3671641167, Damghan, Iran}
\affiliation{Canadian Quantum Research Center 204-3002 32 Avenue Vernon, British Columbia V1T 2L7 Canada}

\author{Saeed Noori Gashti}
\email{saeed.noorigashti@stu.umz.ac.ir}
\affiliation{Department of Physics, University of Mazandaran, P. O. Box 47416-95447, Babolsar, Iran}

\author{Mohammad Reza Alipour}
\email{mr.alipour@stu.umz.ac.ir}
\affiliation{Department of Physics, University of Mazandaran, P. O. Box 47416-95447, Babolsar, Iran}

\preprint{}
\begin{abstract}
We study the thermodynamics of the Taub-NUT AdS black holes by Visser’s holographic method using AdS radius as a constant parameter under the restricted phase space approach. Instead of the variables $P$ and $V$, we deal with the central charge and also chemical potential as a unique couple of conjugate thermodynamic variables. We study some interesting properties of the Taub-NUT black holes \textit{e.g.} supercritical phase equilibrium in the $T-S$ processes, $Q-\Phi$ and the Hawking-Page phase transition in the $\mu-C$ processes. As a consequence, the consistency of the weak gravity conjecture of multi-charge for Taub-NUT-AdS black holes at critical points is proved. We conclude that in the presence of the central charge and assuming $C=\frac{3}{8}\ell^2$, weak gravity conjecture could be satisfied in the Taub-NUT-AdS black holes with $\hat{n}\ll 1$.\\\\
$\dagger$: \textit{Corresponding Author}
\end{abstract}
\maketitle

\section{Introduction}\label{into}

Studying the thermodynamics of black holes was initially motivated and developed by Hawking and Bekenstein \cite{Bekenstein:1972tm,Bekenstein:1973ur,Bardeen:1973gs,Hawking:1975vcx} by working on the thermal properties of black holes from viewpoint of the thermodynamics physics. In such studies, surface gravity and the event horizon area are supposed to be proportional to temperature and entropy. Also, the mass of a black hole is usually considered internal energy. Despite the successes of the mentioned approach, there are still some ambiguities around the issue such as the nature of microscopic degrees of freedom contributing to the entropy of black holes. The holography principle and also AdS/CFT duality play an important role in studying several quantities of black holes in which an AdS black hole state in bulk is equivalent to a thermal state equivalent to dual field theory \cite{tHooft:1993dmi,Susskind:1994vu,Maldacena:1997re,Pahlavani:2013rz}. Besides the above formalisms, one can analyze the thermodynamics of the black holes using different symmetries of a gravitational model \cite{Wald:1993nt}. As an advanced study of AdS black holes, Hawking and Page showed phase transition between AdS black holes with radiation and thermal AdS \cite{Hawking:1982dh}. Additionally, the thermodynamics quantities of black holes concerning the extended phase space thermodynamics are discussed in \cite{Kastor:2009wy,Dolan:2010ha,Dolan:2011jm,Dolan:2011xt,Kubiznak:2012wp,Cai:2013qga,Kubiznak:2016qmn,Xu:2013zea,Xu:2014kwa,Zhang:2017lhl}. In this approach, the thermodynamics behavior of some gravitational models for different black holes is studied by considering a negative cosmological constant proportional to the pressure. Studying such structures provides an excellent incentive to include other parameters of the considered models as novel thermodynamic parameters \cite{Cai:2013qga,Kubiznak:2016qmn,Xu:2013zea,Yang:2020iat,Cong:2021fnf}. Moreover, using $P$, $V$ variables navigates us to the fact that AdS black holes can be assumed as heat engines \cite{Johnson:2014yja,Xu:2017ahm,Ciambelli:2020qny}.

Recently, a new formalism for studying the thermodynamics of black holes has been proposed by including the parameters central charge and the chemical potential as a new couple of conjugate thermodynamics variables \cite{Visser:2021eqk}. Also, in the dual theory, the square number of colors is connected to the central charge \cite{Karch:2015rpa,Maity:2015ida,Wei:2017icx,Rafiee:2021hyj}. Here, we focus on the Taub-NUT-AdS black hole in the restricted phase space (RPS) and study the thermodynamic properties of the model in the presence of the central charge. As the key point of the present work, the weak gravity conjecture (WGC) for such a model can be easily proved at the critical points. 

The layout of the paper is the following. In Section \ref{s2}, we introduce the basic concepts of the Taub-NUT-AdS black hole. In Section \ref{s3}, we examine the thermodynamics properties of the Taub-NUT-AdS black hole in the RPS formalism by  $T-S$  processes $Q-\Phi$. Also, we study the Hawking-Page phase transition in the $\mu-C$ processes in order to prove WGC at critical points. In Section \ref{wgc}, we present a proof of WGC. Conclusions are drawn in Section \ref{s5}.
\section{Taub-NUT-AdS black holes}\label{s2}
We begin with the action of the Taub-NUT-AdS black holes in the framework of Einstein-Maxwell \cite{Jiang:2019yzs,Chen:2019uhp}
\begin{equation}\label{eq1}
\mathcal{S}=\frac{1}{16\pi}\int d^4x\sqrt{-g}(R-2\Lambda-F_{\mu \nu}F^{\mu \nu}),
\end{equation}
where $g$ and $R$ are the determinant of the metric tensor $g_{\mu\nu}$ and the Ricci scalar, respectively. Also, $\Lambda=-\frac{3}{\ell^2}$ is cosmological constant where $\ell$ denotes the anti–de Sitter (AdS) radius. Here, the electromagnetic strength is given by $\mathbf{F}=F_{\mu \nu}=\partial_{\mu}A_{\nu}-\partial_{\nu}A_{\mu}$. By varying the action with respect to the metric $g_{\mu\nu}$, the Einstein equation takes the form
\begin{equation}\label{eq2}
R_{\mu \nu}-\frac{1}{2}Rg_{\mu \nu}+\Lambda g_{\mu \nu}=T_{\mu \nu} ,\hspace{1cm}dG_{\mu \nu}=0,
\end{equation}
where the energy-momentum tensor of the Maxwell field is introduced by
\begin{equation}\label{eq3}
T_{\mu \nu}=F_{\mu \sigma}F_{\nu}^{\sigma}+G_{\mu \sigma}G_{\nu}^{\sigma}    ,\hspace{1cm} \mathbf{G}=\ast\mathbf{F}.
\end{equation}
Also, the electric and magnetic charges inside a two-dimensional surface $S^{2}$ are obtained as \cite{Jiang:2019yzs,Chen:2019uhp}
\begin{equation}\label{eq4}
q_e=\frac{1}{4\pi}\int_{S^2}\mathbf{G},\hspace{1cm}   q_m=\frac{1}{4\pi}\int_{S^2}\mathbf{F}.
\end{equation}
We define metric of the Taub-NUT-AdS black holes as \cite{Alonso-Alberca:2000zeh,Johnson:2014pwa}
\begin{eqnarray}\label{eq5}
ds^2 &=&-f(r)(dt+2n\cos\theta d\varphi)^2+\frac{dr^2}{f(r)}+(r^2+n^2)(d\theta^2+\sin^2\theta d\varphi^2), \nonumber\\
A_{\mu}&=&-[h(r)-h_0] dt+2n h(r) \cos{ \theta} d\varphi,
\end{eqnarray}
where
\begin{equation}\label{eq6}
f(r)=\frac{r^2-2MGr-n^2+4G n^2 g^2+G e^2}{r^2+n^2}-\frac{3n^4-6n^2r^2-r^4}{\ell^2(r^2+n^2)},\hspace{1cm}h(r)=\frac{e r+g(r^2-n^2)}{r^2+n^2},
\end{equation}
where $n$, $M$ and $h_0$ are the NUT parameter, mass of the black hole and an arbitrary constant, respectively. Also, $e$ and $\textrm{g}$ depict degree freedom of the electromagnetic field. Using eq.\eqref{eq4}, the electric and magnetic charges of a black hole on a spherical surface with radius $r$ are calculated as \cite{Jiang:2019yzs,Chen:2019uhp}
\begin{equation}\label{eq8}
q_e(r)=\frac{e(r^2-n^2)-4\textrm{g}rn^2}{r^2+n^2},\hspace{1cm}   q_m(r)=\frac{2n[e r+g(r^2-n^2)]}{r^2+n^2}.
\end{equation}
Asymptotically, The electric and magnetic charges in the limit of $r\rightarrow \infty$ recovers $q_e(r)=e$ and $q_m(r)=2\textrm{g}n$, respectively. The horizon $r=r_H$ is as follows,
\begin{equation}\label{eq9}
Q_m=q_e(r_H),\hspace{1cm}   Q_e=q_m(r_H).
\end{equation}
Also, the entropy at the event horizon takes the following form
\begin{equation}\label{eq10}
S=\frac{A}{4G}=\frac{\pi}{G}(r_H^2+n^2).
\end{equation}
\section{RPS thermodynamics in Taub-NUT-AdS black holes}\label{s3}
In the RPS formalism, the variables pressure $p$ and volume $v$ are not used because of changing their meanings in the holographic viewpoint. On the other hand, two quantities $C$ and $\mu$ play a an important role in  the context of conformal field theory (CFT). Since the central charge $C$ determines the number of degrees of microscopic freedom in CFT, its conjugate $\mu$ is considered as a chemical potential \cite{Gao:2021xtt}. In the context of extended phase space thermodynamics, studying the thermodynamics of the black holes is carried out by using a variable cosmological constnat. As a result, the corresponding dynamical equations are changed. To escape from this, we introduce the thermodynamics of restricted phase space in which Newton's constant is assumed as a variable instead of the cosmological constant and consequently the field equations are kept unchanged. Also, the relationship between the central charge and two cosmological and Newton's constants are introduced as $C=\frac{\ell^{2}}{G}$, where $\ell^{2}=\frac{-3}{\Lambda}$ \cite{Visser:2021eqk,Gao:2021xtt,Zeyuan:2021uol}.

The first law of thermodynamics for a Taub-NUT-AdS black hole in RPS formalism is written as
\begin{equation}\label{eq11}
dM=T dS+\tilde{\phi}_e d\tilde{Q}_e+\tilde{\phi}_m d\tilde{Q}_m+\mu dC.
\end{equation}
This is observed with an Euler-like relation,
\begin{equation}\label{eq12}
M=T S+\tilde{\phi}_e \tilde{Q}_e+\tilde{\phi}_m \tilde{Q}_m+\mu C,
\end{equation}
where $M$, $T$ and $S$ are mass, temperature and entropy of our black hole, respectively. Also, $\tilde{\phi}_e, \tilde{Q}_e$ are the properly re-scaled electric potential and electric charge while $\tilde{\phi}_m, \tilde{Q}_m$ are the properly re-scaled magnetic potential. Generally, the electric and magnetic charges can be defined by the quantities in dual CFT as
\begin{equation}\label{eq13}
\tilde{Q}_e=\frac{Q_e \ell}{\sqrt{G}} ,\hspace{1cm}\tilde{Q}_m=\frac{Q_m \ell}{\sqrt{G}},\hspace{1cm}
\tilde{\phi}_e=\frac{\phi_e\sqrt{G}}{\ell},\hspace{1cm}\tilde{\phi}_m=\frac{\phi_m\sqrt{G}}{\ell}.
\end{equation}
Due to the importance of the mass of a black hole, we consider it as a function of $S, Q_e, Q_m C$. Hence, using equation \eqref{eq10} and the definition $G=\frac{\ell^{2}}{C}$, we rewrite the event horizon radius $(r_H)$ in terms of $S$ by
\begin{equation}\label{eq14}
r_H=\sqrt{\frac{\ell^2 S}{\pi C}-n^2}.
\end{equation}
Then, by using eqs. \eqref{eq8}, \eqref{eq9} and \eqref{eq14}, we express $G$, $e$ and $g$ as
\begin{equation}\label{eq15}
\begin{split}
&G=\frac{\ell^2}{C},\hspace{1cm} e=\frac{-2 \pi n^2 Q_eC +Q_e \ell^2 S+2n Q_mC\sqrt{-n^2\pi^2+\frac{\ell^2 S\pi}{C}}}{\ell^2 s},    \\
&\textrm{g}=\frac{-2 \pi n^2 Q_m C +Q_m \ell^2 S-2n Q_e C\sqrt{-n^2\pi^2+\frac{\ell^2 S\pi}{C}}}{2n \ell^2 s}.
\end{split}
\end{equation}
Setting $f(r_H)=0$ and also using the eqs. \eqref{eq6}, \eqref{eq13}, \eqref{eq14} and \eqref{eq15}, we have the mass of the black hole as
\begin{equation}\label{eq16}
M=\frac{-2C^2n^2 \pi^2(\ell^2+4n^2)+\ell^4 S^2+ \ell^2 \pi[\ell^2 \pi(\tilde{Q}_e^2+\tilde{Q}_m^2)+C S(\ell^2+4n^2)]}{2 C\ell^4 \pi^{\frac{3}{2}}\sqrt{-n^2\pi+\frac{\ell^2 S}{C}}}.
\end{equation}
Also, from eqs. \eqref{eq11} and \eqref{eq16}, we obtain the variables $T,\mu,\tilde{\phi}_e,\tilde{\phi}_m$ as follows
\begin{equation}\label{eq17}
T=\left(\frac{\partial M}{\partial S}\right)_{\tilde{Q}_e,\tilde{Q}_m,C}=\frac{\ell^2\sqrt{-n^2\pi+\frac{\ell^2 S}{C}}[S(\pi C+3S)-\pi^2(\tilde{Q}_e^2+\tilde{Q}_m^2)]}{4\pi^{\frac{3}{2}}( \pi n^2 C -\ell^2 S)^2},
\end{equation}
\begin{equation}\label{eq18}
\tilde{\phi}_e=\left(\frac{\partial M}{\partial \tilde{Q}_e}\right)_{S,\tilde{Q}_m,C}=\frac{ \tilde{Q}_e}{C\sqrt{-n^2+\frac{\ell^2 S}{\pi C}}},
\end{equation}
\begin{equation}\label{eq19}
\tilde{\phi}_m=\left(\frac{\partial M}{\partial \tilde{Q}_m}\right)_{S,\tilde{Q}_e,C}=\frac{ \tilde{Q}_m}{C\sqrt{-n^2+\frac{\ell^2 S}{\pi C}}},
\end{equation}
\begin{eqnarray}\label{eq20}
\mu=\left(\frac{\partial M}{\partial C}\right)_{S,\tilde{Q}_e,\tilde{Q}_m}&=&
\frac{(\ell^2+4n^2)(-4 \pi^3 C^3n^4+6\pi^2 C^2\ell^2 n^2 S)+\ell^6S[\pi^2(\tilde{Q}_e^2+\tilde{Q}_m^2)+S^2]}
{4C^2 \ell^4 \pi^\frac{3}{2}(\pi n^2 C-\ell^2S)\sqrt{-n^2\pi+\frac{\ell^2 S}{C}}} -\nonumber\\&&            -\frac{C \ell^4 \pi[2n^2\pi^2(\tilde{Q}_e^2+\tilde{Q}_m^2)+(\ell^2+6n^2)S^2]}{4C^2 \ell^4 \pi^\frac{3}{2}(\pi n^2 C-\ell^2S)\sqrt{-n^2\pi+\frac{\ell^2 S}{C}}}.
\end{eqnarray}
By re-scaling $S$, $\tilde{Q}_e$ and $\tilde{Q}_m,C$ as $S \rightarrow \alpha S$, $\tilde{Q}_e\rightarrow \alpha \tilde{Q}_e$, $\tilde{Q}_m \rightarrow \alpha \tilde{Q}_m$, $C \rightarrow \alpha C $, the mass of the black hole $M$ \eqref{eq16} is re-scaled as $\rightarrow\alpha M$ while the parameters $T$, $\mu$, $\tilde{\phi}_e$ and $\tilde{\phi}_m$ remain unchanged. Consequently, $M$ is first-order homogeneity and $T$, $\mu$, $\tilde{\phi}_e$ and $\tilde{\phi}_m$ are zero-order homogeneity which are defined as extensive and intensive variables, respectively.

In the following, we investigate the thermodynamic processes \textit{e.g.} $T-S$, $\tilde{\phi}_e-\tilde{Q}_e$, $\tilde{\phi}_m-\tilde{Q}_m$ and $\mu -C$ considering only one pair of conjugate intensive-extensive variables.
\subsection{$T-S$ processes }
The $T-S$ curve at fixed $\tilde{Q}_e,\tilde{Q}_m, C$ is a thermodynamic characteristic including a first-order phase transition that becomes second-order at the critical point. To obtain the critical point in the $T-S$ curve at fixed $\tilde{Q}_e$, $\tilde{Q}_m$ and $C$, we use 
\begin{equation}\label{eq21}
\left(\frac{\partial T}{\partial S}\right)_{\tilde{Q}_e,\tilde{Q}_m,C}=0, \hspace{1cm} \left(\frac{\partial^2 T}{\partial S^2}\right)_{\tilde{Q}_e,\tilde{Q}_m,C}=0.
\end{equation}
Using eqs.\eqref{eq17} and \eqref{eq21} for the critical parameters, we have  
\begin{equation}\label{eq22}
S_c=\frac{\pi C( \ell^2 +12 n^2)}{6 \ell^2},\hspace{1cm} \tilde{Q}_{e,c}^2+\tilde{Q}_{m,c}^2=\frac{C^2}{36}\left(1+48\frac{n^2}{\ell^2}+144\frac{n^4}{\ell^4}\right).
\end{equation}
Also critical values of $T$ and $M$ are obtained by
\begin{equation}\label{eq23}
T_c=\sqrt{\frac{2}{3}}\frac{\sqrt{\ell^2+6n^2}}{\pi \ell^2}, \hspace{1cm}   M_c=\frac{1}{3}\sqrt{\frac{2}{3}}\frac{C(\ell^2+6n^2)^{\frac{3}{2}}}{\ell^4}.
\end{equation}
In the limit of $n \rightarrow 0$ and $\tilde{Q}_{m,c} \rightarrow 0$, eqs. \eqref{eq22} and \eqref{eq23} are reduced to $T_c=\sqrt{\frac{2}{3}}\frac{1}{\pi \ell}$, $S_c=\frac{\pi C}{6 }$ and $\tilde{Q}_{e,c}=\frac{C}{6}$ which present the Reissner-Nordström black holes in the RPS formalism \cite{Maldacena:1997re}.

To obtain the thermodynamic potential of the black holes using the Euclidean action, we usually encounter the redshift degeneracy in the action. As a solution to remove this shortcoming, we attempt to consider a suitable counterterm (coming from the AdS/CFT correspondence) in the Euclidean action. Consequently, we can solve the modified Euclidean action in order to find the thermodynamic potential associated with the canonical and grand canonical ensembles. As an example, we refer to the Gibbs energy potential $G=\frac{i}{\beta}$ where $i$ and $\beta$ are the Euclidean action and the inverse of Hawking temperature, respectively. Then, we can find other thermodynamic quantities such as  $S=-\frac{dG}{dt}$, $J=-\frac{dG}{d\Omega}$ and $Q=-\frac{dG}{d\phi}$ using the obtained the Gibbs energy potential. This approach is equivalent to the method using the first law of thermodynamics to find the thermodynamic quantities.

Also, we can rewrite the thermodynamic quantities in terms of the Helmholtz free energy and its conjugates as $S=-\frac{dF}{dt}$, $J=-\frac{dF}{d\Omega}$ and $Q=-\frac{dF}{d\phi}$ since it connects to the Euclidean action as $F=\frac{i}{\beta}$. Analogous to the Gibbs energy, the result coming from the Helmholtz free energy is still compatible with the result coming from the first law of thermodynamics.

In the RPS formalism, Newton's constant as a variable is situated out of the Euclidean action so that it does not add any extra terms to the counterterm. Hence, one can find different values of counterterm for different black holes in order to remove the redshift degeneracy \cite{Cvetic:2001bk,Dayyani:2016gaa}.

Now, we start with the Helmholtz free energy as
\begin{equation}\label{eq24}
F(T,\tilde{Q}_e,\tilde{Q}_m,C)=M(S,\tilde{Q}_e,\tilde{Q}_m,C)-TS.
\end{equation}
From eqs. \eqref{eq22},\eqref{eq23} and \eqref{eq24}, the critical point of Helmholtz energy takes the following form
\begin{equation}\label{eq25}
F_c=\frac{C}{3\sqrt{6}}\frac{\sqrt{\ell^2+6n^2}}{\ell^2}.
\end{equation}
For our purposes, it might be worth introducing the parameters
\begin{equation}\label{eq26}
s=\frac{S}{S_c},   \hspace{1cm}  t=\frac{T}{T_c}, \hspace{1cm} q^2=\frac{\tilde{Q}_e^2+\tilde{Q}_m^2}{\tilde{Q}_{e,c}^2+\tilde{Q}_{m,c}^2}, \hspace{1cm} f=\frac{F}{F_c}.
\end{equation}
Using eq.\eqref{eq26} and the definition $\hat{n}=\frac{n}{\ell}$, one can rewrite eqs.\eqref{eq17} and \eqref{eq24} as 
\begin{equation}\label{eq27}
t=\frac{-144\hat{n}^4(q^2-3s^2)+24 \hat{n}^2(-2q^2+3s^2+3s)+(-q^2+3s^2+6s)}{8\sqrt{\hat{n}^2+6}[ s+6\hat{n}^2(-1+2 s)]^{\frac{3}{2}}},
\end{equation}
and
\begin{eqnarray}\label{eq28}
f&=&
\frac{144\hat{n}^4(-2+q^2+2s+s^2)+24\hat{n}^2(-3+2q^2+4s+s^2)+(\textrm{q}^2+s^2+6s)}{4\sqrt{1+6\hat{n}^2}\sqrt{ s+6\hat{n}^2(-1+2s)}}-\left(1+12\hat{n}^2\right)st.
\end{eqnarray}
\begin{figure*}[!hbtp]
	\centering
	\includegraphics[width=.45\textwidth,keepaspectratio]{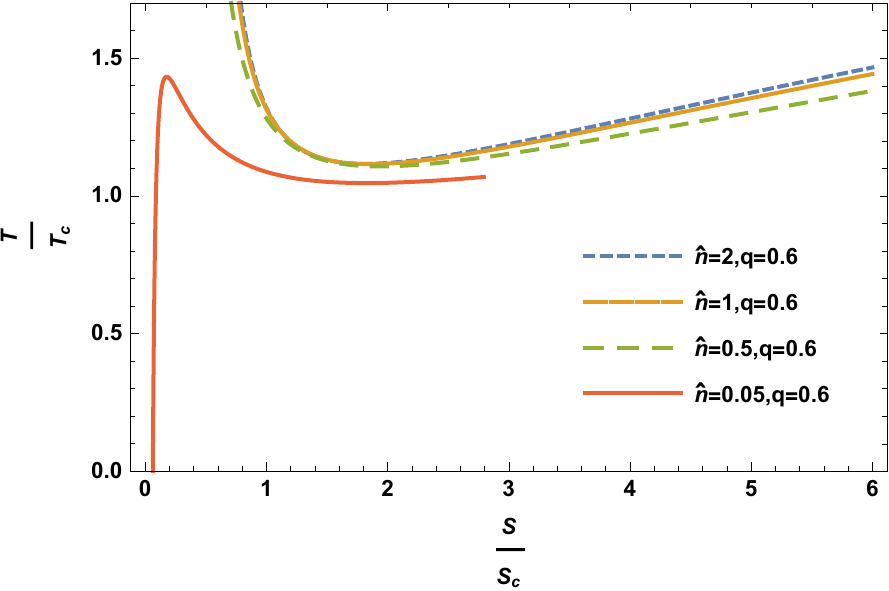}\hspace{0.1cm}
	\includegraphics[width=.45\textwidth,keepaspectratio]{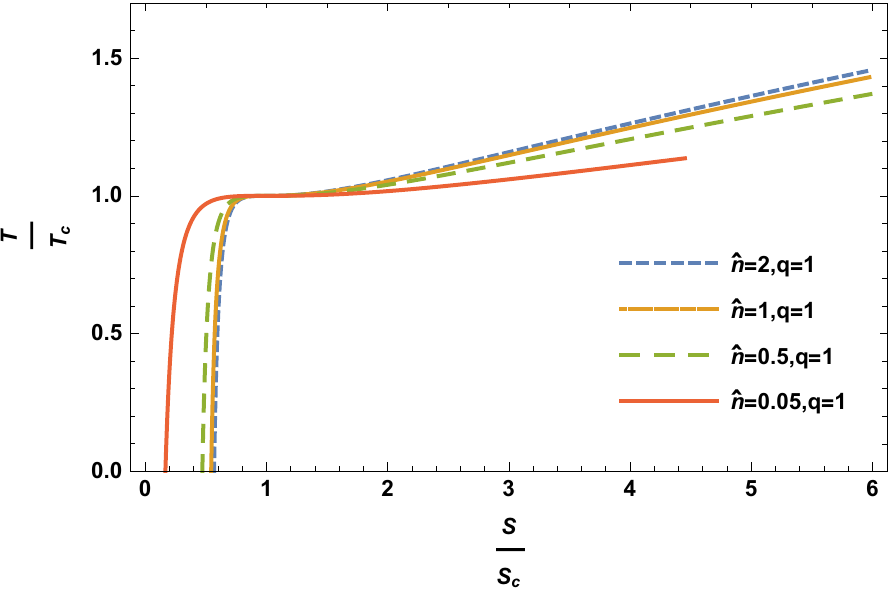}\hspace{0.1cm}
	\includegraphics[width=.45\textwidth,keepaspectratio]{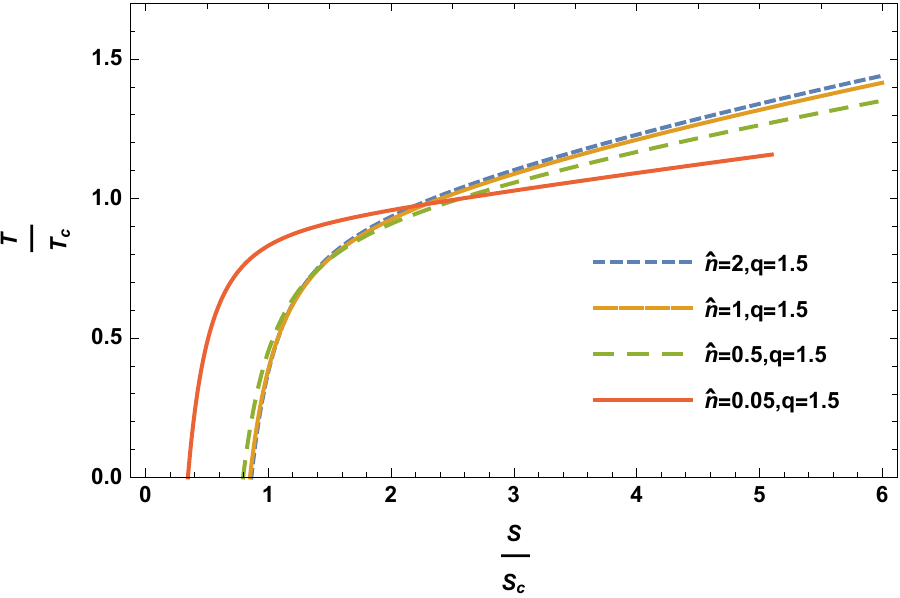}
	\caption{The critical behavior of ($T/T_c -S/S_c$) with respect to constant parameters of Taub NUT black hole. Notice that for all values of q and smaller values of $\hat{n}$, we encounter the critical point in order to proof the WGC.}
	\label{fig1}
\end{figure*}
Fig.\ref{fig1} presents the critical behavior of $T/T_c -S/S_c$ in the Taub NUT black hole for $\hat{n} = 0.05, 0.5, 1, 2$ with different $q=0.6, 1, 1.5$, separately. From the figure, we find that for $q<1$, there is a critical point in $\hat{n}=0.05$. Also, $\hat{n}\leq 0.13$ is more suitable for the Taub-NUT-AdS black holes at the critical point in RPS formalism. Also, in the limit of $\hat{n}\rightarrow0$, the diagrams recover the results in \cite{Maldacena:1997re,Wald:1993nt}. We note that both constant parameters $q$ and $\hat{n}$ play an essential role in the drawn curves so that for smaller values, there is a scaled problem for the red curve with no affection on the results (see the top-left panel of Fig.\ref{fig1}). Moreover, the curve does not show any other maximum or minimum points. This is also found through the proof of the WGC in which the values of $\hat{n}$ must be much smaller than 1. In order to establish an acceptable expression for the WGC, and as we have shown in eqs. (\ref{eq32}) to (\ref{eq35}), we are required to introduce some particular values of the critical points within certain limits.
\subsection{$\mu-C$ processes }
\begin{figure*}[!hbtp]
	\centering
	\includegraphics[width=.45\textwidth,keepaspectratio]{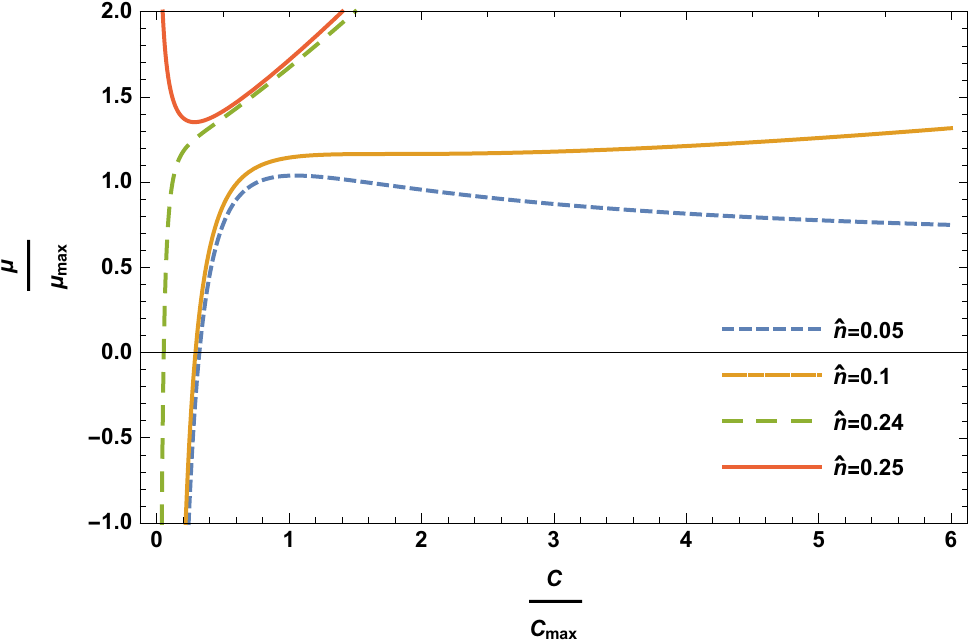}
	\caption{The behavior of ($\mu/\mu_{max} -C/C_{max}$) with respect to constant parameters of Taub NUT black hole.}
	\label{fig2}
\end{figure*}

To study the $\mu-C$ processes, we obtain the maximum $\mu$ and $C$ concerning eq.\eqref{eq20} for the $\hat{n}\ll 1$. So, we have approximately 
\begin{eqnarray}\label{eq29}
C_{max} & \approx & \frac{3S(\pi^2 q^2+S^2)}{\pi(8n^2\pi^2+S^2+12n^2S^2)},\nonumber\\
\mu_{max} & \approx  &
\frac{[S^2+4\hat{n}^2(2\pi^2 q^2+3S^2)][S^2-4\hat{n}^2(5\pi^2 q^2+3s^2)]}{6\sqrt{3} \ell S^4}\sqrt{\frac{S^2+\hat{n}^2(5\pi^2 q^2+9S^2)}{\pi^2 q^2+S^2}}.
\end{eqnarray}
Here, we introduce the dimensionless parameters
\begin{equation}\label{eq30}
c=\frac{C}{C_{max}},   \hspace{1cm}  m=\frac{\mu}{\mu_{max}}.
\end{equation}
Then, the re-scaled $\mu-C$ connection is obtained as
\begin{equation}\label{eq31}
m=\frac{(3c-1)+2n^2(27c^2-15 c+8)}{2c^{\frac{3}{2}}[1-n^2(3 c-8)]}.
\end{equation}
In here for simplicity of the discussion we define the equation
\begin{equation}
\Omega=\frac{-1+10\hat{n}^2+\sqrt{1+4\hat{n}^2-284\hat{n}^4}}{36\hat{n}^2}.    
\end{equation}
In Fig.\ref{fig2}, we show the behavior of $\mu/\mu_{max} -C/C_{max}$ with respect to the constant parameters of the Taub NUT black hole in $\hat{n}=0.05, 0.1, 0.24, 0.25$. In such a case, we face with three cases: i) For $C=\Omega C_{max}$, $\mu$ becomes zero. ii) For $C>\Omega C_{max}$ in $\hat{n}\leq 0.24$, the free energy becomes zero so that the microscopic degrees of freedom are repulsive. iii) For $C<\Omega C_{max}$, the degrees of freedom are attractive. 
Also, in $\hat{n}=0$, we reproduce the results of the RN-AdS and Kerr-AdS black holes \cite{Gao:2021xtt,Zeyuan:2021uol}.
\section{Weak Gravity Conjecture perspective}\label{wgc}
WGC as a swampland program has recently been challenged by different frameworks \cite{Arkani-Hamed:2006emk,Vafa:2005ui,Ooguri:2006in,Arkani-Hamed:2006emk,Sadeghi:2020ntn,Sadeghi:2022wgx}. More precisely, the existence of an infinite tower of exactly stable states in a fixed direction in charge space is prohibited if we encounter a UV complete model of quantum gravity. Arguments against such an infinite tower include a topic that may lead to a species problem, etc \cite{Susskind:1995da}. It is generally consistent with all well-known explicit evidence of string compactifications and other conjectures of quantum gravity such as the finiteness principle and the absence of global symmetries \cite{Susskind:1995da}. These equivalences can be considered to fit the conjectures, such as statements about the (in-) stability of asymptotically large extremal black holes. In quantum gravity, we expect that the initial states with super-Planckian masses will appear to distant observers as a series of black hole solutions in the framework of the low-energy effective field theory \cite{Susskind:1993ws,Horowitz:1996nw}. One of the most famous predictions of WGC is associated with a single charge while here in the Taub-NUT-AdS black holes, we face the electric and magnetic charges. Hence, we require to redefine the WGC condition as \cite{Cheung:2014vva}
\begin{equation}\label{eq32}
\frac{q_e^2+q_m^2}{M^2}>1,
\end{equation}
where $q_e,q_m $ and $M $ are  electric , magnetic charges and mass of the black hole, respectively . Considering the critical points, the relations \eqref{eq22} and \eqref{eq23}, as well as the scaling of the relation \eqref{eq13}, $\hat{n}=\frac{n}{\ell}$ and $\tilde{M}_c =\sqrt{\frac{3}{8}} M \ell$, we obtain 
\begin{equation}\label{eq33}
\frac{\tilde{Q}_{e,c}^2+\tilde{Q}_{m,c}^2}{\tilde{M}_c^2}=\frac{1+48 \hat{n}^2+144\hat{n}^4}{(1+6\hat{n}^2)^3}.
\end{equation}
Clearly, for $\hat{n}>1$, the WGC condition in eq.\eqref{eq33} is not valid while for $\hat{n}\ll 1$, the relation \eqref{eq33} is given as 
\begin{equation}\label{eq34}
\frac{\tilde{Q}_{e,c}^2+\tilde{Q}_{m,c}^2}{\tilde{M}_c^2}=1+30 \hat{n}^2+ \mathcal{O}(\hat{n}).
\end{equation}
As a secondary proof of WGC, we rewrite eq. \eqref{eq33} using eqs.\eqref{eq13} and \eqref{eq15}, as follows (without considering the scale change)
\begin{equation}\label{eq35}
\frac{Q_{e,c}^2+Q_{m,c}^2}{M_c^2}=\frac{3}{8}\frac{\ell^2}{C}\left(\frac{1+48 \hat{n}^2+144\hat{n}^4}{(1+6\hat{n}^2)^3}\right).
\end{equation}
Now assuming $C=\frac{3}{8}\ell^2$, WGC is proved for the Taub-NUT-AdS black hole with respect to $\hat{n}\ll 1$. 

\section{Conclusions}\label{s5}
In this paper, we investigated the Taub-NUT-AdS black hole thermodynamics from a new method with a particular structure that is founded on the thermodynamics of Visser’s holographic. The AdS radius is considered a constant, called the restricted phase space thermodynamics (RPST). Instead of the variable ($P$,$V$), this method supposed a new variable, viz central charge, and chemical potential, as a unique couple of conjugate thermodynamic variables. The Euler relation keeps in this formalism automatically. Also, the mass homogeneity in the first order and the intensive variables homogeneity in the zeroth-order are constructed explicitly and shown earlier in this formalism
Therefore, we tried to investigate the thermodynamics of the Taub-NUT-AdS black hole in the RPS formalism and obtained its features such as supercritical phase equilibrium in the $T-S$ , $Q-\Phi$ processes and Hawking-Page phase transition in the $\mu-C$ processes.
Then we benefited from this information and proved WGC of multi-charge for Taub-NUT-AdS black hole in this mentioned phase space at critical points.
In this study, we considered two different states, the first without central charge, in which case the WGC condition holds for $\hat{n}\ll 1$. And the second state, we considered the central charge with assuming  $C=\frac{3}{8}\ell^2$, the WGC is proved for the Taub-NUT-AdS black hole with  respect to $\hat{n}\ll 1$.

As an outlook for our future works, we can search for a possible universal value of the central charge due to testing the method for other black holes. Moreover, one can explore for stronger proof to create a relationship between the WGC and the AdS/CFT correspondence when some modifications of the conjectures are assumed. In forthcoming papers, these aspects will be deeply investigated.\\

\noindent\textbf{Data Availability Statement:}\\
The authors confirm that the data supporting the findings of this study are available within the article and its supplementary material. Raw data that support the findings of this study are available form the corresponding author, upon reasonable request. 
\bibliographystyle{ieeetr}
\bibliography{biblo}
\end{document}